# Contextualising and Aligning Security Metrics and Business Objectives: a GQM-based Methodology

Eleni Philippou, Sylvain Frey, Awais Rashid

**Abstract**—Pre-defined security metrics suffer from the problem of contextualisation, i.e. a lack of adaptability to particular organisational contexts – domain, technical infrastructure, stakeholders, business process, etc. Adapting metrics to an organisational context is essential (1) for the metrics to align with business requirements (2) for decision makers to maintain relevant security goals based on measurements from the field. In this paper we propose SYMBIOSIS, a methodology that defines a goal elicitation and refinement process mapping business objectives to security measurement goals via the use of systematic templates that capture relevant context elements (business goals, purpose, stakeholders, system scope). The novel contribution of SYMBIOSIS is the well-defined process, which enforces that (1) metrics align with business objectives via a top-down derivation that refines top-level business objectives to a manageable granularity (2) the impact of metrics on business objectives is explicitly traced via a bottom-up feedback mechanism, allowing an incremental approach where feedback from metrics influences business goals, and vice-versa. In this paper, we discuss the findings from applying SYMBIOSIS to three case studies of known security incidents. Our analysis shows how the aforementioned pitfalls of security metrics development processes affected the outcome of these high-profile security incidents and how SYMBIOSIS addresses such issues.

## 1 INTRODUCTION

Security metrics play a fundamental role in security decision-making and relative prioritisation of security requirements within an organisation. Critically, if security metrics can effectively capture the realisation of business objectives, this can ensure that security fits with organisational goals. On the other hand, such alignment of security metrics with business objectives can allow decision makers to evolve and update objectives based on relevant measurements in the field. Reference information security metrics sources such as NIST 800-55 [24] and ISO 27004 [15] focus on the measurement process itself and propose to reuse pre-defined, supposedly universal, security metrics. Such approaches suffer from two key limitations that are addressed by SYMBIOSIS (see Fig. 1):

1) **Lack of contextualisation:** Capturing the relevant technical and organisational factors that influence security in a particular organisation is critical in order to design relevant measurements. In the absence of a clear design methodology, supposedly universal metrics are hard to adapt effectively to particular organisations, domains and contexts [8]. A well-defined contextualisation is also necessary to trace and continuously adapt metrics to ever-evolving security landscapes, organisational constraints and business contexts. A particular challenge is that of capturing the interplay between a (possibly complex) technical infrastructure and a (possibly complex) organisation: a misunderstanding of who does what in such a socio-technical context can hide critical design failures.

2) **Lack of alignment between security metrics and business objectives**: Lack of contextualisation makes it difficult to align security metrics and business objectives. On the one hand, security metrics must align with business objectives, in a top-down fashion, in order to take into account the specific risk factors (cost, impact on resources, business continuity) of the organisation. On the other hand, feedback from security measurements has to be carried bottom-up for strategic business decisions to better integrate concrete security risks and situations from the field. Such a process must be traceable, repeatable and incremental, in order for an organisation to adapt continuously to a changing threat landscape, as well as its own business evolution.

Goal-driven approaches to security measurements such as [14], [16], [23] do promote a more structured way of designing information security metrics and discuss the alignment of business objectives and security goals. However, they lack a systematic definition of goals and measurements, such as comprehensive templates to capture various facets of the goals and measurements, nor do they provide a detailed derivation methodology. This lack of a systematic approach limits the ability to evaluate the effectiveness of the resulting metrics with respect to business objectives in a

• E. Philippou and S. Frey conducted the work while at Lancaster University, UK. S. Frey is currently at Google DeepMind. A. Rashid is with the University of Bristol, UK.
E-mail: eleni.p.philippou@gmail.com, frey.sylvain@gmail.com, awais.rashid@bristol.ac.uk

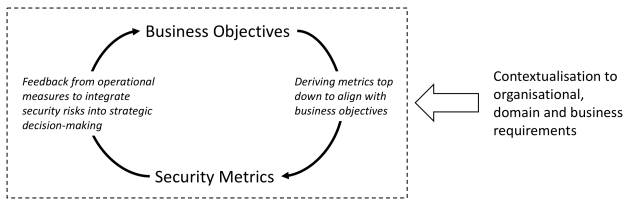

Fig. 1. Gaps in state-of-the-art and novel contributions of SYMBIOSIS.

reproducible way.

Approaches addressing such concerns have previously been designed for, and applied with great success to the Software Engineering industry, guiding the creation of goal-driven metrics for various projects. The most prominent amongst those approaches include the Goal-Question-Metric (GQM) approach [2]. GQM and its successors – GQ(I)M [20] and GQM+ [1] – provide a well-defined methodology to contextualise the creation of metrics, e.g. for assessing software quality and team productivity.

Establishing *security* metrics raises specific concerns however. Security goals are often orthogonal – and sometimes contradictory – to business concerns: confidentiality concerns requiring to reduce the availability of an asset, for instance. In addition, security is a crosscutting concern that overlaps usual organisational structures and requires the collaboration of different, potentially conflicting, stakeholders: Chief Information Security Officer (CISO), engineers, managers, policy makers, decision makers, third parties, etc [9]. Capturing this organisational dimension of security is a challenge [27], and as a result, aligning security measurements with business objectives is still difficult.

In this paper, we propose SYMBIOSIS (SecuritY Metrics and BusIness ObjectiveS, Integrated and Synchronised), a methodology that promotes the integration of business goals, organisational context and security risks into the security metrics development process. SYMBIOSIS builds on the foundations of GQM and provides security-specific templates and methodological elements to contextualise security metrics and align them with business objectives. The novel contributions of SYMBIOSIS are as follows (cf. Fig. 1):

- A goal elicitation and refinement process that defines business objectives and refines them progressively, down to a manageable granularity where concrete security measurement goals and metrics can be established. The process uses systematic templates to contextualise the business objectives, measurement goals and metrics with respect to the technical and organisational specificities of the case at hand.
- A well-defined measurement process that ensures the traceability of metrics back to security and business objectives, explicating their impact and facilitating their integration and co-evolution with business processes, via an incremental approach.

SYMBIOSIS is the first methodology to cover security from the highest level of decision making to the concrete practice of measurements while encompassing both technical and organisational aspects. It integrates several orthogonal dimensions usually not considered together: reg-ulatory compliance, privacy policies, privacy control and enforcement, and operational considerations such as "how to enforce the policy" and "what to do when a breach is detected". Such a comprehensive approach enables the alignment of security metrics with a business perspective. Such breadth of scope is complemented by SYMBIOSIS's adaptability, which allows to tailor the methodology to a wide variety of cases, at small and large scales. It should be noted, however, that SYMBIOSIS limits itself to modelling business objectives, measurement goals and metrics, the relation between them and the methodology to establish such models. SYMBIOSIS does not, for the time being, provide an explicit decision-making framework for driving security strategies.

In the following, we discuss background and related work on security metrics and GQM (Section 2). We then introduce SYMBIOSIS, demonstrating it in detail through application on a case study, along with a thorough discussion of its strengths and limitations (Section 3). We present and discuss findings from applying SYMBIOSIS on three comprehensive case studies, demonstrating how the aforementioned pitfalls of security metrics development processes affected the outcome of several high-profile security incidents and how SYMBIOSIS addresses such issues (Section 4). Finally, we conclude the paper by summarising the contributions of SYMBIOSIS and identifying directions for future work (Section 5).

## 2 BACKGROUND & RELATED WORK

We review related work on metrics with respect to two evaluation criteria (summarised in Table 1):

- The **contextualisation** of security metrics with respect to technical, organisational and security factors, and whether or not systematic templates are used to define such metrics.
- The **alignment** between business objectives and security metrics, including derivation of metrics from objectives, carrying feedback from metrics to business objectives, and whether or not systematic templates are used to define goals.

### 2.1 Reference Sources for Information Security Metrics

Reference sources such as NIST 800-55 [24], ISO 27004 [15], CISWG [6], CIS Security Metrics [5] or securitymetrics.org [3] adopt a catalogue approach: they present various reference metrics classified into categories and documented with scenarios and examples. The intended audience is expected to find and reuse metrics matching their needs, however, little support is provided for finding the particular use-cases that apply to the situation at hand.

Without a well-defined adaptation methodology (in the case of CISWG [6], CIS Security Metrics [5] and security-metrics.org [3]), the contextualisation of metrics relies on arbitrary examples and use cases, which limits their expressiveness and precludes alignment with business objectives. When a contextualisation methodology is available, it either does not refer to business objectives, as is the case with ISO 27004 [15], or only promotes alignment with regulations and laws, for instance, NIST 800-55 [24].

TABLE 1
Positioning of SYMBIOSIS with respect to related work.

| | Standard Information Security Metrics [24], [15], [6], [5], [3] | Goal-Driven Approaches to Security [14], [16], [13], [23] | Goal-Oriented Requirements Engineering [26], [25], [10] | GQM Methodologies [2], [20], [1] | SYMBIOSIS |
|---|---|---|---|---|---|
| **Contextualisation** • technical factors (scope, dependencies, domain) | | partial | ✓ | ✓ | ✓ |
| • organisational factors (scope, stakeholders) | | partial | ✓ | ✓ | ✓ |
| • security factors | ✓ | ✓ | partial | | ✓ |
| • metric templates | ✓ | | | | ✓ |
| **Alignment** • well-defined derivation from business goals to metrics | | partial | | ✓ | ✓ |
| • well-defined feedback from metrics to business goals | | | | ✓ | ✓ |
| • goal templates | | | ✓ | ✓ | ✓ |

## 2.2 Goal-driven approaches to information security

[14], [16], [13], [23] promote a more structured way of designing information security measurements that takes stakeholders and higher-level goals into account. However, these approaches do not provide systematic templates for goals, metrics and measurements and do not go deeper than suggesting using a well-defined, step-by-step goal elicitation process that would be tailored to security. As a result, the adaptability and traceability supported by these methods is still limited.

## 2.3 Goal-oriented requirements engineering

Contextualisation has been a core concern of the requirements engineering community. Goal languages such as i* [26] and KAOS [25] have introduced comprehensive, extensible models of stakeholders, systems, and the interplay between the two, including detailed goal templates and taxonomies. These approaches are not meant to be applied out of the box to security settings. More specific frameworks have been developed for particular use cases, see, for instance, [10] for privacy compliance.

Although the focus of goal-oriented requirement engineering approaches is on producing software requirements instead of metrics, they highlight important considerations when modelling technical systems and the surrounding organisation, such as system inter-dependencies or the relationship between stakeholders of different natures and their business objectives.

## 2.4 Goal Question Metric Methodologies

The Goal Question Metric (GQM) methodology is a reference for goal-driven measurement in software engineering. The original formulation by Basili and Weiss [2] defined the foundations of the method as a succession of steps: establish goals, formulate questions from the goals, design and perform data measurements based on the questions. As such, this generic basis provides little more than "formulate questions" to support the contextualisation of measurements. In addition, the initial "goals" use no particular template and make no reference to business objectives.

Later evolutions of GQM refined the methodology in order to address its shortcomings. Goal Question Indicator Measurement (GQ(I)M) [20] refined the early goal establishing steps by explicitly referring to top business goals and adding goal refinement and subgoal derivation steps. The resulting indicators and measurements are, therefore, explicitly aligned with business objectives. This alignment is top-down only however: GQ(I)M provides little indication as to how feedback from measurements could influence business objectives or the priorities between them. In addition, GQ(I)M does not provide well-defined templates but relies mainly on natural language questions to drive the process, which also limits its traceability.

GQM+ Strategies [1] is another extension of GQM that focuses on alignment with business objectives. The method uses templates to describe business objectives in terms of object, context and relation with other goals. This degree of formalism makes the approach more traceable and reusable. It does not encompass security however, and therefore does not contextualise the particular aspects of security goals and their interactions with business goals, namely, how security goals often conflict with business goals (e.g., how ensuring the confidentiality of data may require to reduce its availability), and how security concerns crosscut organisations and affect the (potentially conflicting) viewpoints of different stakeholders (e.g., CISO, policy makers and third parties). Finally, unlike goals which have well-defined templates, metrics are not captured in a systematic fashion. This leaves a number of open questions regarding the applicability of the method to security metrics, for instance: What is the metric's measurement method? What is the interpretation of the measurement? Which stakeholders are to be informed of the results, and with what frequency?

## 2.5 Summary and positioning

The state of the art is summarised in Table 1:

- Standard Information Security metrics and goal-driven approaches for security do not or very partially address the issue of alignment. Furthermore, contextualisation of technical and organisational factors is limited within such approaches.

- Goal-oriented requirements engineering and GQM approaches do address the issues of contextualisation and alignment, however, these approaches are not specific to security and its particular trade-offs (e.g., business goals and security goals conflicting with each other).

As shown in Table 1, existing approaches do not address these challenges fully. A comprehensive framework is required that addresses the problems of contextualisation and alignment in a holistic fashion. Note that the problem cannot simply be addressed by combining arbitrary features from multiple approaches. As shown in Fig. 1, a systematic approach is needed whereby business objectives are stepwise refined into security metrics but traceability is maintained from metrics back to the business objectives as the organisation and its systems (and security controls) evolve. SYMBIOSIS, discussed next, aims to address these issues.

## 3 THE SYMBIOSIS APPROACH

### 3.1 Overview

The SYMBIOSIS methodology consists of 4 main steps, shown in Figure 2:

1) Define business objectives and refine them recursively into sub-objectives of a finer granularity.
2) When the refinement of business objectives reaches a manageable granularity, derive security measurement goals that capture the achievement of such objectives.
3) From security measurement goals, derive security metrics via questions (in a GQM fashion).
4) Once security metrics are established, feedback results to the related business objectives and update the whole model accordingly.

The approach follows the overall structure of a GQM methodology (business objective decomposition via templates and use of questions to derive metrics) enriched with specific security elements, namely templates for security measurement goals and security metrics. These are inspired from standard Information Security metrics and goal-driven approaches to security which do provide security-specific models and templates yet lack the structure and methodological rigour of GQM (cf. Table 1). While presenting the methodology, we will draw the reader's attention towards the following novel contributions of SYMBIOSIS:

- The ability to capture the complexity of a technical system, of an organisation, and of the interplay between the two.
- The ability to capture orthogonal dimensions of the situation, including regulatory compliance, privacy policies, privacy control and enforcement, and operational considerations when facing a security event.
- The ability to adapt the granularity of the modelling from the global strategic goals of a large organisation down to individual systems and stakeholders.

### 3.2 Sample use-case

For the purposes of demonstration, we focus on the case of JP Morgan, one of the case studies considered in this work [11], [18], [12], [17]. Just like any financial institution in the US, JP Morgan has a continuing obligation under the Gramm-Leach-Bliley Act of 1999 "to respect the privacy of its customers and to protect the security and confidentiality of those customers' nonpublic personal information". In 2014, JP Morgan was compromised by hackers, stealing account holder contact information (including names, addresses, phone numbers and email addresses) for 76 million households and 7 million small businesses. The following sections use this particular use case to illustrate the use of SYMBIOSIS in an organisation.

### 3.3 Step 1: Define and derive business objectives

*Step 1.1: Define top-level business objectives*

The first step of SYMBIOSIS is to identify the business objectives that drive the organisation's efforts. These high-level objectives will constitute the underlying motivation for the entire measurement process and are essential to the successful implementation and maintenance of a measurement program.

For the purpose of successfully eliciting business objectives, a team-setting whereby the measurement team will have the opportunity to interview and discuss with higher-level managers and strategic decision makers is advisable. Organisation-wide documentation is a key starting point for the discussion, including for instance:

- Findings of recent assessments (threat landscape, vulnerabilities, risks, compliance).
- Policies and documented procedures.
- Relevant regulatory and legislative provisions the organisation should take into consideration.

The outcome of this step should be a set of formalised business objectives that are prioritised according to certain criteria defined by the measurement team and management. For the purpose of formalising business objectives, we propose the use of a contextualisation template, shown in Table 2. Once this template is completed, a business objective can be formulated in natural language as follows :

*One of our primary business objectives is to <achieve purpose> with respect to <object> within that <scope>, from the viewpoint of <viewpoint> while taking into account <context and limitations>. Achieving this business objective will impact/affect/depend on <other relevant business objectives>.*

*Step 1.2: Derive and refine sub-objectives recursively*

Having formally defined and prioritised business objectives, the next step in the process involves the identification of potential strategies for their achievement and the selection of the one that best manages to take the different constraints and limitations identified into account. As in the first step, the engagement and participation of members of the higher-management and strategic decision makers in the process of defining the high-level strategy is of vital importance to the process. In attempting to identify potential strategies, a useful statement to have in mind would be:

```
1: Define and derive business objectives
      → 1.1: Define top-level business objectives
      → 1.2: Derive and refine sub-objectives recursively

2: Define security measurement goals

3: Derive security metrics via security measurement questions
      → 3.1: Derive security measurement questions and consider answer sources
      → 3.2: Derive security metrics

4: Utilise security metrics
      → 4.1: Undertake security measurements
      → 4.2: Provide business feedback from security measurements [Return to 1.2]
      → 4.3: Adapt metrics to changing business objectives
```

Fig. 2. Overview of the SYMBIOSIS methodology.

TABLE 2
Formalised Business Objective

| | |
|---|---|
| **Identifier** | Unique identifier for the business objective. |
| **Object** | The systems and organisations the objective focuses on. |
| **Scope** | The systems and organisations that affect or are affected by the objective. |
| **Purpose** | What is to be achieved with respect to the object within the scope. |
| **Viewpoint** | Which stakeholders are primarily interested in the achievement of the objective. |
| **Context / Limitations** | Contingencies that have to be taken under consideration when planning on how to achieve the purpose (including cost and budgeting constraints, regulatory/legislative requirements, time constraints, personnel availability, etc.). |
| **Relationship with other objectives** | What other business objectives may affect or be affected by the objective. |

"To achieve <business objective> while taking into account <context and limitations> we could . . . ".

Once a strategy has been decided upon, its steps have to be clearly identified and the decision has to be documented and fully justified. The process then continues by identifying sub-goals associated with each step of the strategy. These sub-goals essentially capture various aspects of the higher-level business objective and aim to decompose it into more manageable pieces.

Considering that the strategy was clearly and comprehensively defined in the previous step, a good indication of the stakeholders who could be of more value to the process at this step should be available. These stakeholders are the people responsible for the achievement of different aspects of the higher-level strategy; their view of the goals they need to achieve in order to succeed in their area of responsibility will help move the process closer to meaningful measurement goals.

For each of the areas of the strategy, relevant stakeholders should be prompted to identify their objectives, which are essentially sub-objectives of the higher level business objective formalised at the beginning of this process, and formalise them. For the purposes of assisting this process, the template shown in Table 2 can be reused. Once the template has been completed, the sub-objectives can be formalised as follows:

*Assess the <object> including all elements within <scope>, for the purpose of <purpose> from the viewpoint of <viewpoint>. When doing so, <constraints and limitations> should be taken into account. This objective is expected to impact/affect/depend on <other relevant business sub-objectives and objectives>*

Where multiple sub-objectives are identified for a single strategy area, these should be prioritised according to criteria defined by stakeholders and the prioritisation along with its justification documented. This key step defines important trade-offs inherent to security contexts, for instance:

- confidentiality vs. availability of user data
- intrusiveness of monitoring systems vs. employee privacy
- security investments vs. acceptable risks and expected losses

These trade-offs are captured by the "relationship with other goals" field of the template shown in Table 2 and the corresponding decision can be traced back to the stakeholders involved in the definition of the objective, also captured in the template.

Once business sub-objectives are identified and formalised, the relevant stakeholders should consider how they would go about achieving those sub-objectives. This process will result in a new set of steps for achieving the sub-objective and will bring us closer to realising what would make a meaningful measurement.

**Note:** Decomposing strategies to the point where measurement goals can be refined is the first major milestone of our methodology. It should be exercised with caution so as to avoid an unmanageably large set of sub-strategies and subgoals. The purpose of the refinements are to bring us closer to a point where meaningful measurement goals can be derived. As soon as a strong indication of what could be measured to guide and inform our process for achieving an objective exists, the refinement process should terminate and the process should proceed to the next step.

*Example*

We illustrate step 1 in the methodology via an application to the JP Morgan case. We establish top-level business objectives, then derive sub-objectives down to a manageable granularity, demonstrating Symbiosis's ability to capture orthogonal dimensions of security (here, regulatory compliance, security policies and business objectives), including the complexity of the underlying organisation, in an adaptable fashion. Figure 3 provides an indication of what could constitute a top-level Formalised Business Objective Template in this case. This top-level business objective, BO1, can be formulated in natural language as follows:

*BO1: One of our primary business objectives is to apply a systematic approach to effectively manage security of our information security assets company-wide, from the viewpoint of the CEO and CISO, while taking into account the legally imposed deadline for doing so. This business objective depends on the achievement of BO1.1*

Having defined and formalised the business objective, the measurement team and strategic decision makers need to consider possible strategies towards its achievement. For the purposes of our example, we assume that upon considering various alternatives, the organisation decided to go with ISO27001/2[1], since it provides the only international benchmark for information security management verified by an independent audit, and a potential accreditation could serve as a competitive advantage. An example strategy is shown in Figure 3, which breaks down BO1 into more manageable chunks: implement an Information Security Management System, or ISMS (step 1), audit it for compliance (step 2), assess its effectiveness (step 3).

For each one of the three steps comprising the strategy for BO1, a number of sub-objectives and strategies for their achievement can be defined. We focus now on the sub-objective corresponding to the evaluation of the implemented ISMS (step 3). This sub-objective (see Figure 3 is formulated as follows :

*BO1.1: Analyse the implemented ISMS including all elements within its scope such as policies, procedures, control objectives, controls(...) for the purpose of assessing their effectiveness from the viewpoint of the CISO, before the next scheduled audit, within the allotted budget. This objective depends on BO1.1.1*

Again, a strategy must be designed to implement objective BO1.1. We assume here that the stakeholders decide to follow the decomposition in "controls" defined within the ISO27001/2 standard (e.g., "A5" for Information Security Policies, "A7" for Human Resource Security, "A9" for Access Control, etc.): each of these "controls" will be assessed individually. It should be noted that such a strategy, illustrated in Figure 3, is still at a high level of abstraction: the definition of the "effectiveness" of a security control and the way it should be assessed is left for further refinements, namely when measurement objectives and metrics themselves are defined.

The last step of the refinement phase that we describe here involves the definition of a narrower business objective corresponding only to an aspect of the strategy (specifically the control objective "A7 – Human Resource Security" of ISO27001/2) and the strategy associated with its achievement (see Figure 3):

*BO1.1.1: Analyse the controls implemented for the purposes of ensuring Human Resource Security, including all control relevant to human resource security prior, during and following the termination or change of employment, for the purpose of assessing their effectiveness from the viewpoint of the Information Security Operations Manager, before the next scheduled audit, within the allotted budget.*

This business objective has now a manageable granularity from which a meaningful security measurement goal – i.e., the meaning of the "effectiveness" of security controls and how to "assess" it – can be derived, described and illustrated as discussed next.

### 3.4 Step 2: Define security measurement goals

Having defined our strategies, the next step in the process involves the determination and definition of security measurement goals. Unlike business objectives whose aim is to express a high-level corporate vision, security measurement goals aim to help the organisation determine what needs to be known for the purpose of deciding on the success or failure of a particular strategy.

For the formalisation of security measurement goals, we propose the use of the template presented in Table 3. This template addresses various aspects that we have identified as critical for the successful definition of security metrics. Once this template is completed, a security measurement goal can be formalised as follows:

*Analyse <object> for the purpose of <purpose> <focus> of all elements within <scope> with respect to <criteria> from the viewpoint of <stakeholder> in the context of <environment>. This measurement goal is expected to impact/affect/depend upon <relevant goals and/or objectives>.*

TABLE 3
Formalised Security Measurement Goal

| Identifier | Unique identifier for the security measurement goal. |
|---|---|
| Object | What is to be measured. |
| Purpose | Why does the measurement take place. |
| Focus | What attributes of the object are of interest. |
| Scope | What does measuring the object entail. |
| Criteria | In terms of which the purpose and focus should be established. |
| Viewpoint | For whom is the measurement taken. |
| Context | What should one take into account when planning or interpreting the measurement. |
| Relationship with other goals | How the measurement affects/impacts/depends upon other security measurement goals. |

*Example*
In the JP Morgan case, the strategy for BO1.1.1 (Figure 3) is decomposed enough for meaningful security measurement goals to be derived. For the purpose of our example, we

---

1. JP Morgan was ISO27k-compliant at the time of the breach.

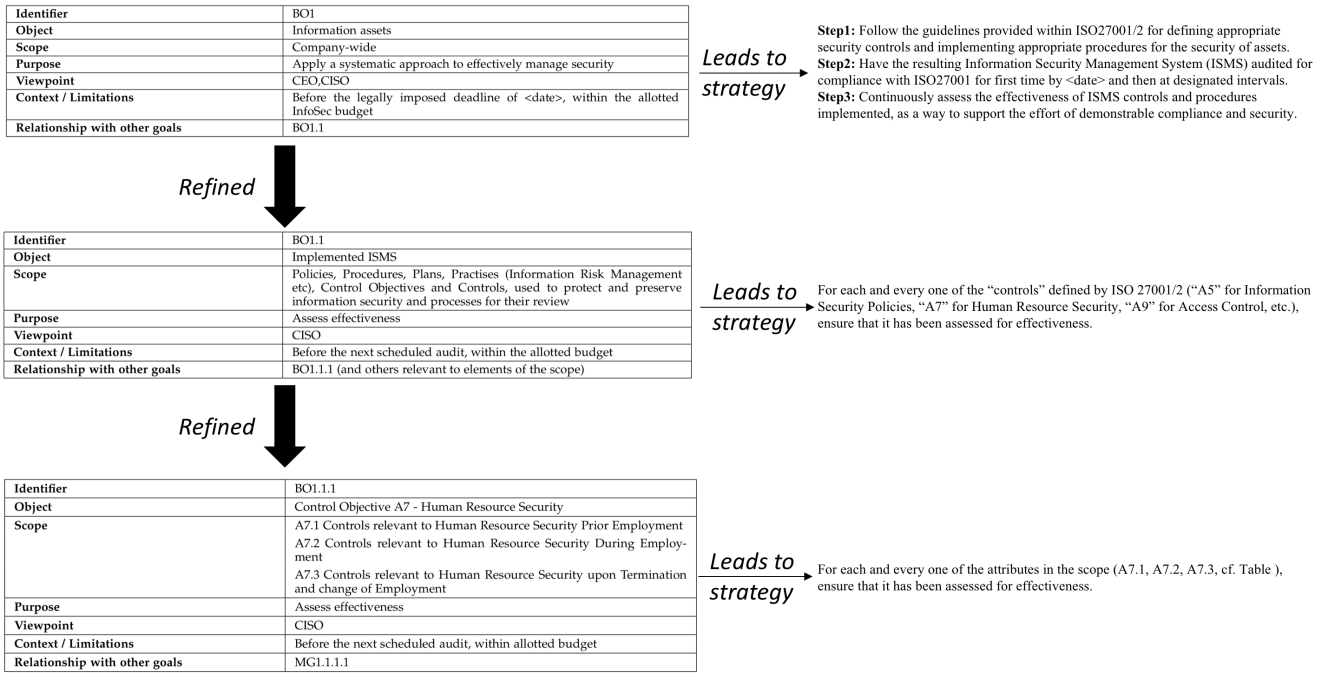

Fig. 3. Example of Step 1 of the Symbiosis methodology.

illustrate one such goal that aims to assess the effectiveness of the information security awareness, education and training process, part of control A7.2 in the scope of BO1.1.1. Table 4 illustrates the formalisation of the security measurement goal that can be formulated as follows :

*MG 1.1.1.1 Analyse the information security awareness, education and training process and specifically the content and activities, for the purpose of evaluating their effectiveness, with respect to currentness, reviewing frequency (...), from the viewpoint of the manager responsible for security awareness, education and training taking into account the timing (before the next audit) and risk considerations to define priorities.*

### 3.5 Step 3: Derive security metrics via security measurement questions

*Step 3.1: Derive security measurement questions and consider answer sources*

With the security measurement goals defined and formalised, the next step in the process involves the derivation of security measurement questions and the consideration of potential information that could help us answering those questions. As security measurement questions, we define questions whose answers help us determine whether a security measurement goal is being achieved or help demonstrate substantive progress towards its achievement.

These questions could be of various natures according to the security measurement goal to which they correspond and the criteria for its achievement. Examples include (in this case, considering a given business objective $BO_i$):

- How do we measure the scope of personnel and systems concerned by security objective $BO_i$?

- How do we assess whether the reviewing frequency for objective $BO_i$ is appropriate?
- How do we make sure security objective $BO_i$ is kept aligned with the current threat landscape and risk assessment?
- How do we assess the relevance of objective $BO_i$ with respect to current practices and business processes?
- How do security events and incidents affect objective $BO_i$, and vice versa?

These questions are at the crux of a security-focused contextualisation and alignment of metrics and business objectives: the very contribution of Symbiosis. Derived questions and elements that could lead to answering those questions are of key importance, and they should be documented by the measurement team, before proceeding to the next step. Security measurement questions serve as guidelines for eliciting the kind of information that will need to be collected through measurement.

In attempting to provide answers to questions, potential attributes of the object of interest and how these could be measured to meaningfully answer questions is a key step towards the metrics to follow. The results of the measurement of such attributes are what we refer to as "base measurements" – not dissimilarly to ISO27004. Examples of such base measurements could include:

- number of security events, attack detections, incidents, successful defences, etc.
- number of secure assets, vulnerable assets, compromised assets, etc.
- proportion of breaches / successful defences / etc. in detected incidents.

TABLE 4
Example of Formalised Measurement Goal

| Identifier | MG1.1.1.1 |
|---|---|
| Object | Security Awareness, Education and Training process |
| Purpose | Evaluating |
| Focus | Effectiveness |
| Scope | Security Awareness training content, Security Awareness and Training activities |
| Criteria | Currentness/ Reviewing frequency/ Contents of the delivered training and how these map to organisational and standard-regulatory requirements/ Tailoring to person's roles, responsibilities and risks/ Delivered at appropriate frequency/ Attended by all relevant audience/ Frequency of refresher activities - But also in terms of observed incidents/ observed events that could be traced back to the lack of training/awareness/education |
| Viewpoint | Responsible Manager for Security Awareness, Education and Training process |
| Context/Limitations | Before the next scheduled audit, taking into account risk considerations to define priorities |
| Relationship with other goals | BO1.1.1 |

- proportion of secure / vulnerable / compromised / etc. assets.
- coverage of risk assessment, security tests, penetration testing, etc.
- dates / frequency / average period for scheduled maintenance, vulnerability detection and security patch, incident response, security training refresher, etc.

Any kind of attribute that, following its measurement, produces a result that when reasonably combined with others can produce a metric which meaningfully answers a question is considered to be a valid base measurement.

*Step 3.2: Derive security metrics*

With security measurement questions clearly defined and considerations around data-sources and base measurements already in the measurement team's mind, the next step in the process involves the derivation of security metrics. In its simplest form, a security metric describes how various data – used for answering questions such as the ones described above – can be combined in a meaningful way that demonstrates substantive progress towards the achievement of a security measurement goal.

Defining such security metrics requires defining combinations of data that could meaningfully answer questions and serve security measurement goals. "Meaningful" here means that these data capture one of, or the relationships between, the following aspects of the organisation (non-exhaustive list):

- Technical infrastructure and human organisation.
- Business processes and practices.
- General threat landscape and known specific threats.
- Known vulnerabilities.
- Risk profile and risk appetite.
- Security policies and practices.
- Past security events and security exercises.

The set of potential security metrics derived at this step should be checked against the security measurement goal and questions to verify their ability to indicate progress towards the achievement of the goal. This is the second milestone of our methodology.

Once a conclusive decision is reached regarding the best combinations of base security measurements, these combinations should be documented along with information that is of importance for the purposes of ensuring the repeatability of the measurements and the validation of the derived security metrics. Specifically, for each security metric definition finalised, a template such as the one introduced in Table 5 should be completed.

TABLE 5
Formalised Security Metric

| Identifier | Unique identifier for the security metric. |
|---|---|
| Description | Including dates of creation, last modification and last review. |
| Goal of measurement | The measurement goal associated with this metric. |
| Base measurements | The raw numbers that are to be collected. |
| Measurement method | How the base measurement shall be carried out. |
| Measurement function | Formal definition of how base measurements are to be processed. |
| Measurement interpretation | How the measurement should be interpreted in the domain's terms. |
| Reporting method and frequency | How often should the measurement be performed and reported. |
| Stakeholders | Who performs the measurement and who it is reported to. |

This template extends the approach of ISO27004 with the business objective alignment dimension of SYMBIOSIS. The template aims to ensure that no relevant information with respect to a security metric has been left behind, defining systematically what a derived metric means, what objectives and stakeholders it is associated with, or what question it is aiming to answer.

Among other information, the template defines the security measurement goal(s) a security metric is associated with, the kind of base measurements it combines, how these were collected, by whom, using what sources, how these were combined, how the measurement should be interpreted, how it should be reported, to whom and how often.

**Note:** It is important to note that in the process of trying to define a security metric, the need for a refinement of security measurement questions or measurement goals might arise. In those cases, the refinement should take place before proceeding to formalising a metric. A security metric should only be formalised if it is considered to be final and sufficient for the purposes of answering the question and bringing us closer to achieving the goal.

*Example*

Having formalised measurement goal MG1.1.1.1 regarding Security Awareness, Education and Training Process, a range of questions need to be derived. A potential outcome of this exercise could resemble that presented in Table 6. The table demonstrates the transition from Questions to Security Metrics through the consideration of Base Measurements. The different questions address the following aspects of measurement goal MG1.1.1.1:

- Q1.1.1.1.1: up-to-dateness of the security training programme.
- Q1.1.1.1.2: reviewing frequency of the security training programme.
- Q1.1.1.1.3: alignment of security training programme with roles, responsibilities and risks.
- Q1.1.1.1.4: coverage of the organisation's personnel by the security training programme.
- Q1.1.1.1.5: frequency of refresher activities.
- Q1.1.1.1.6: influence between known incidents and security training.

The security metrics provided in Table 6 are not yet formalised. In fact, they constitute mere indications of a potentially meaningful combination of base measurements. Prior to formalising a security metric a quick re-examination of whether it actually responds to the question and the measurement goal it was derived from, is required. This, in turn, may mean change to the metrics. Such a back-and-forth workflow is key to ensuring the alignment between metrics and business objectives, and the traceability offered by SYMBIOSIS helps support this in a systematic way.

Figure 4 offers a schematic overview of the steps in the process thus far. Due to space constraints, the explicit association between questions and base-measurements that help answer them and metrics that can be derived by combining these base metrics are presented for only two of the derived metrics. All templates are also omitted from the model for the same reason.

Following this process each metric that is verified to give answers to the questions and helps establish a progress towards the achievement of the measurement goal is formalised according to the metric template. An example is given in Table 7. It should be noted that the metric captures together a technical infrastructure ("base measurement" attribute and the following 3 in the template), a business strategy ("goal of measurement" attribute), an organisation

("stakeholder" attribute), and offers an operational perspective ("reporting method and frequency").

### 3.6 Step 4: Utilise security metrics

Once the full derivation is established, from top-level business objectives to security metrics, SYMBIOSIS enters its last phase, characterised by continuous measurements and adaptations instead of one-off steps. Several activities take place in parallel:

- Security measurements, based on the established metrics.
- Feedback from security measurements to related business objectives.
- Changes in the business objectives influencing the security metrics.

These activities must be performed continuously, in a back-and-forth fashion between business objectives and security metrics, so as to maintain the alignment between the two.

#### 3.6.1 Step 4.1: Undertake security measurements

The contextualisation and use of templates to describe metrics provide a rich support for carrying out security measurements. The templates provide not only a detailed, explicit measurement method, but also the necessary context to understand and interpret the measurement's results, and to whom to communicate the results. The latter point is key whether the results are positive or not: a bad security measurement should be as relevant as a good security measurement, and communicated to the relevant stakeholders in the organisation.

*Example*

The team in charge of carrying out measurements for metric ME 1.1.1.1.1 (cf. Table 7) performs a monthly collection of the base measurement, as specified by the metric. It usually gets an excellent score, superior to 91%, which does not require further action but is logged and forwarded to the corresponding manager responsible for Awareness, Education and Training. When an unexpected low score (70%) is recorded, action can be taken immediately and in proportion: the manager is notified that closer attention may be required, while the different stakeholders involved (trainers, new hires) are readily identifiable to investigate the reasons behind the bad score (logging mistake? lack of monitoring of new hires? overloaded training programme?). In case the issue increases in criticality (score below 60%, according to the metric), escalation measures are straightforward: the CISO, main stakeholder for the corresponding business objective (BO1.1.1, cf. Figure 3), can be notified of the issue, and the business consequences are explicit, in terms of the affected objectives (BO1.1 and BO1).

#### 3.6.2 Step 4.2: Provide Business feedback from security measurements

From the point of view of decision makers, contextualised measurements provide precious decision support. Security measurement results (good or bad) can be situated in a

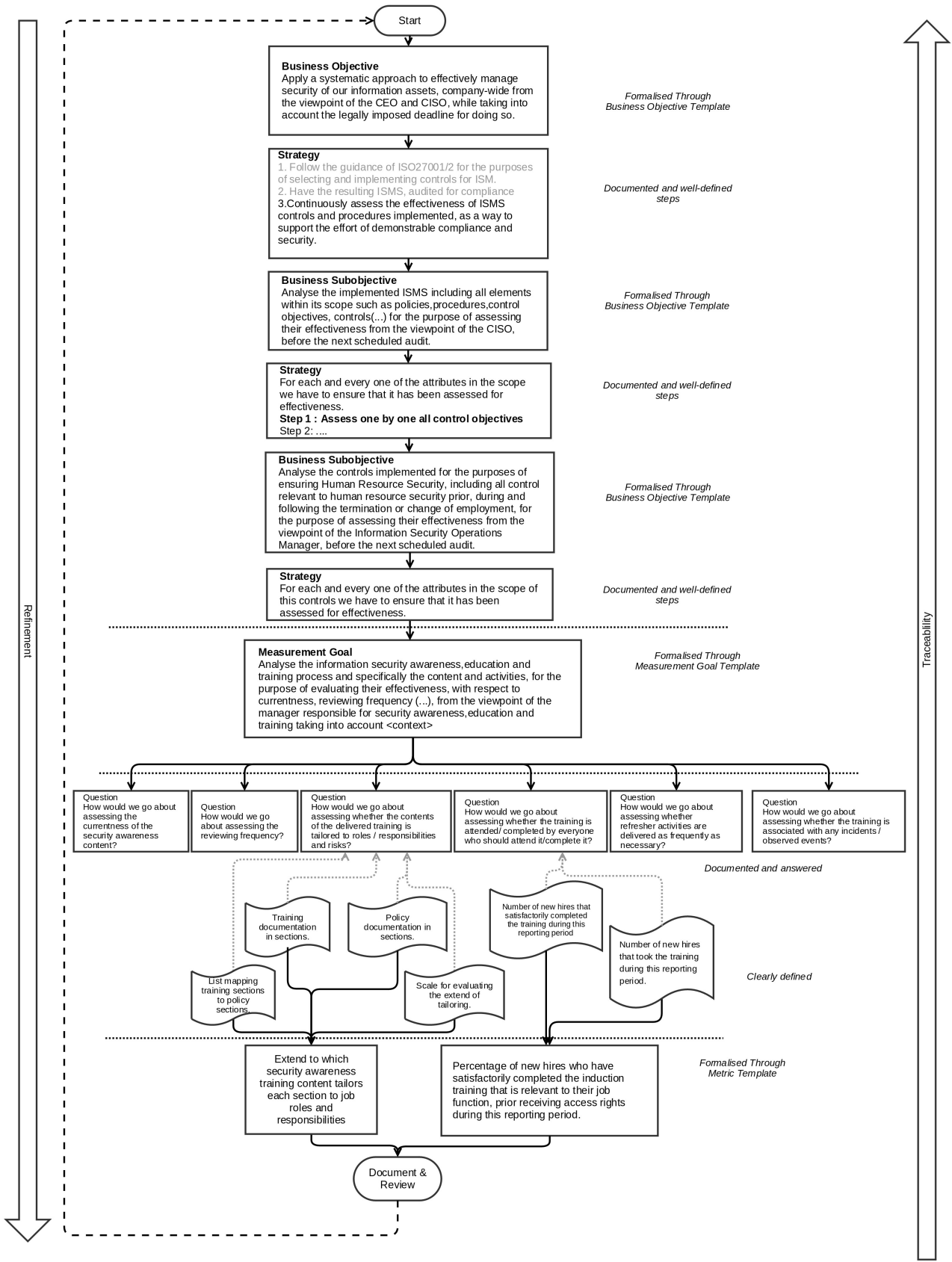

**Start**

**Business Objective**
Apply a systematic approach to effectively manage security of our information assets, company-wide from the viewpoint of CEO and CISO, while taking into account the legally imposed deadline for doing so.

*Formalised Through Business Objective Template*

**Strategy**
1. Follow the guidance of ISO27001/2 for the purposes of selecting and implementing controls for ISM.
2. Have the resulting ISMS, audited for compliance
3. Continuously assess the effectiveness of ISMS controls and procedures implemented, as a way to support the effort of demonstrable compliance and security.

*Documented and well-defined steps*

**Business Subobjective**
Analyse the implemented ISMS including all elements within its scope such as policies,procedures,control objectives, controls(...) for the purpose of assessing their effectiveness from the viewpoint of the CISO, before the next scheduled audit.

*Formalised Through Business Objective Template*

**Strategy**
For each and every one of the attributes in the scope we have to ensure that it has been assessed for effectiveness.
**Step 1 : Assess one by one all control objectives**
Step 2: ....

*Documented and well-defined steps*

**Business Subobjective**
Analyse the controls implemented for the purposes of ensuring Human Resource Security, including all control relevant to human resource security prior, during and following the termination or change of employment, for the purpose of assessing their effectiveness from the viewpoint of the Information Security Operations Manager, before the next scheduled audit.

*Formalised Through Business Objective Template*

**Strategy**
For each and every one of the attributes in the scope of this controls we have to ensure that it has been assessed for effectiveness.

*Documented and well-defined steps*

**Measurement Goal**
Analyse the information security awareness,education and training process and specifically the content and activities, for the purpose of evaluating their effectiveness, with respect to currentness, reviewing frequency (...), from the viewpoint of the manager responsible for security awareness,education and training taking into account <context>

*Formalised Through Measurement Goal Template*

**Question**
How would we go about assessing the currentness of the security awareness content?

**Question**
How would we go about assessing the reviewing frequency?

**Question**
How would we go about assessing whether the contents of the delivered training is tailored to roles / responsibilities and risks?

**Question**
How would we go about assessing whether the training is attended/ completed by everyone who should attend it/complete it?

**Question**
How would we go about assessing whether refresher activities are delivered as frequently as necessary?

**Question**
How would we go about assessing whether the training is associated with any incidents / observed events?

*Documented and answered*

Training documentation in sections.

Policy documentation in sections.

Number of new hires that satisfactorily completed the training during this reporting period

Number of new hires that took the training during this reporting period.

List mapping training sections to policy sections.

Scale for evaluating the extend of tailoring.

*Clearly defined*

Extend to which security awareness training content tailors each section to job roles and responsibilities

Percentage of new hires who have satisfactorily completed the induction training that is relevant to their job function, prior receiving access rights during this reporting period.

*Formalised Through Metric Template*

**Document & Review**

Refinement

Traceability

Fig. 4. Example of SYMBIOSIS approach

TABLE 6
Example of Questions, Base Measurements and Security Metrics

| Questions | Base-Measurements | Security Metrics |
|---|---|---|
| Q1.1.1.1 How would we go about assessing the currentness of the security awareness content? (Whereby currentness is defined as the alignment with the provisions of the latest business policy and procedures documented and in line with the requirements of applicable laws and regulations?) | • Number of training sections that have been assessed against the most recent policies and procedures and with respect to the requirements of applicable laws and regulations, to determine if they were aligned.<br>• Total number of sections in training content.<br>• Scale for evaluating the extend of alignment.<br>• Number amongst those which were not deemed "up-to-date".<br>• Number amongst those not deemed "up-to-date" that were scheduled for review.<br>• List mapping training sections to policy sections.<br>• List mapping training sections to relevant regulatory and legislative requirements, wherever possible. | • Percentage of training awareness content that has been assessed for alignment with the most current policies and procedures of the organisation and the requirements of applicable laws and regulations.<br>• Extend to which training awareness content was deemed to be in line with : a)company policy and procedures b) requirements or applicable laws and regulations.during the last alignment evaluation.<br>• Percentage of training awareness content that has not been deemed as "aligned and up-to-date" during the last alignment evaluation, for which updating has been scheduled before the next correctness evaluation.<br>• Percentage of training awareness content for which alignment evaluations have been conducted at defined intervals. |
| Q1.1.1.2 How would we go about assessing the reviewing frequency? | • Date of previous training content reviews.<br>• Date-log for all previous reviews. | • Frequency with which training content is being reviewed to determine its alignment with the most current policies and procedures of the organisation and the requirements of applicable laws and regulations. |
| Q1.1.1.3 How would we go about assessing whether the content of the delivered training is tailored to roles / responsibilities and risks? | • Training documentation in sections.<br>• Policy documentation in sections.<br>• List mapping training sections to policy sections.<br>• Scale for evaluating the extend of tailoring. | • Extend to which security awareness training content tailors each section to job roles and responsibilities. |
| Q1.1.1.4 How would we go about assessing whether the training is attended / completed by everyone who should attend it/ complete it? | • Number of new hires who took the training during this reporting period.<br>• Number of new hires that satisfactorily completed the training during this reporting period.<br>• Total number of new hires during this reporting period.<br>• Number of employees that took the refresher training during this reporting period.<br>• Number of employees that satisfactorily completed the refresher training during this reporting period.<br>• Total of employees during this reporting period. | • Percentage of new hires who have taken the induction training during this reporting period.<br>• Percentage of new hires who have satisfactorily completed the induction training that is relevant to their job function, prior receiving access rights during this reporting period.<br>• Percentage of employees who took their periodic security awareness training refresher during this reporting period.<br>• Percentage of employees who have satisfactorily completed their periodic security awareness training refresher relevant to their job function during this reporting period. |
| Q1.1.1.5 How would we go about assessing whether refresher activities are delivered as frequently as necessary? | • Date of previous refresher training.<br>• Date-log of all previous refresher trainings. | • Frequency with which security awareness refresher training is being delivered. |
| Q1.1.1.6 How would we go about assessing whether the training is associated with any incidents / observed events? | • Number of incidents traced back to human factors in the past reporting period.<br>• Total number of incidents in the past reporting period for which root-cause analysis has been performed. | • Percentage of security incidents for which root cause analysis conducted in the past reporting period determined that root cause was a "human factor" failure. |

TABLE 7
Example of a Formalised Security Metric

| Identifier | ME 1.1.1.1.1 |
|---|---|
| Description | Metric to determine the effectiveness of the process in place for delivering training to new hires in terms of its successful completion. <creation Date/Last reviewed/modified date> |
| Goal of measurement | MG 1.1.1.1 (Q1.1.1.1.4) |
| Base measurement | Number of new hires that took the training during this reporting period. Number of new hires that satisfactorily completed the training during this reporting period. Total number of new hires during this reporting period. |
| Measurement method | Count log entries within this reporting period where the field of new-hire awareness training is designated as "attended"/ the outcome is designated as "successfully completed" and the date is within the reporting period of interest |
| Measurement function | (Number of new hires that satisfactorily completed the training during this reporting period. /Number of new hires that took the training during this reporting period. )* 100 |
| Measurement interpretation | 0-60% Intervention is required - we need to determine what hinders the effectiveness of the process. 61-90% We need to watch the metric closely as it might indicate something needs to be altered 91-100% No need for change. |
| Reporting method and frequency | Monthly collection of base measurements, quarterly analysis and reporting of selected measurements. The use of a bar chart with different colour-coded bars based on the interpretation of the measurement could make for a valuable reporting method |
| Stakeholder | Manager responsible for security Awareness, Education and Training. (and anyone else responsible for communicating this metric higher up the hierarchy). |

global business scheme established during the goal elicitation and derivation process. In case these results do not align with expectations, business objectives can be adapted accordingly to prescribe more realistic goals for the organisation, or new strategies can be adopted to tackle the issue at play. This activity is key notably to deal with dynamic technologies and emerging threat actors that can rapidly change the threat landscape: decision makers need to stay informed to be reactive to the technical and technological context.

*Example*

After a series of low scores measured for metric ME 1.1.1.1.1, the manager responsible for Awareness, Education and Training (main associated stakeholder) investigated and identified that the training team was under-staffed compared to a recent increase in recruitment. Provision for expanding the training team is requested to the higher management, justified with a synthetic history of measurements and the global business consequences of the current situation.

### 3.6.3  Step 4.3: Adapt metrics to changing business objectives

Although not as fast-paced as modern threat landscapes, business objectives are also expected to change during the lifetime of an organisation – be it for pure business reasons or in reaction to changes in the security environment. Having an explicit model of business goals and the traceability to security metrics provides a support for managing such change. In case a business objective is added, removed or changed, the associated objectives are explicitly identified, and the consequences of the change can be anticipated. The consequences in terms of metrics can be worked out incrementally, starting from the previous known situation.

*Example*

Given an increase in the sophistication of social engineering attacks, the CISO of the company decides to harden the protections of the organisation against it. The change in policy is the following: the initial mandatory Security Awareness training for all new employees must now be complemented by a refresher every 6 months. On one hand, the existing resources allocated for the training objective (Business Objective BO1.1.1) give an indicator of the expected investment. On the other hand, the existing measurement goal and derived metric can be adapted to reflect the policy change: the new interpretation of the measurement result – or possibly, a new measurement goal and metric altogether – is established to take the new goal into account, while the increase in logging and reporting activities can be estimated based on the current baseline.

## 4  EVALUATION OF THE APPROACH

This section provides an evaluation of SYMBIOSIS: the strengths and limitations of the approach are discussed and illustrated via concrete examples selected from case studies. In total, we applied SYMBIOSIS to three real-world incidents: JP Morgan, Anthem and Heartland, summarised in Table 8. For each of these, the methodology was applied, starting from a (partial) set of business objectives down to security metrics, in order to identify the key points where the use of SYMBIOSIS would have prevented said incidents from happening.

In section 3 we have presented the details of applying SYMBIOSIS to the JP Morgan case. A similar exercise was conducted for the other two case studies: a selection of the most significant outcomes is presented as an appendix to this paper. In the following sections, we limit ourselves to discussing certain particular aspects of the case studies that highlight the strengths of SYMBIOSIS.

Such a *post-mortem* analysis allows to evaluate the effectiveness of SYMBIOSIS on real-world incidents with organisation of various sizes and purposes (cf. Table 8). Beyond these particular cases, the goal of this section is also to identify common security mistakes gathered from real-world practices, and to illustrate how SYMBIOSIS prevents such mistakes via its systematic approach.

TABLE 8
Summary of the validation case studies.

| Name | References | Year | Organisation type | Organisation size | Summary of the attack |
|------|-----------|------|-------------------|-------------------|----------------------|
| Heartland | [4] | 2008 | Bank | < 4000 employees | Despite regular audits and PCI compliance, an old server was compromised for months and allowed attackers to capture unencrypted traffic on internal bank networks. Tenths of millions of cardholder records were stolen. |
| Anthem | [21] | 2015 | Health insurer | < 40,000 employees | Attackers got access (probably via phishing) to administrative credentials with read access to personal records databases. Nearly 80 million personal records were stolen. |
| JP Morgan | [11], [18], [12], [17] | 2014 | Bank | > 50,000 employees | Attackers infiltrated the corporate network after stealing an employee's credentials (probably via phishing). Months later, a repository containing stolen passwords and usernames was identified by a third party and traced back to JP Morgan, where data exfiltration was still going on. It is estimated that the account holder information for 76 million households and 7 million small businesses were stolen. |

## 4.1 Strengths

### 4.1.1 Breadth

SYMBIOSIS is the first approach to cover security from the highest level of decision making to the concrete practice of measurements while encompassing both technical and organisational aspects. SYMBIOSIS captures key dimensions such as:

- Technical complexity, for instance: hidden system interdependencies leading to cascading failures, lack of isolation allowing escalation of privileges.
- Organisational complexity, for instance: interleaved business processes interfering with good security practices, overlapping stakeholders / chains of authority obscuring security policies.
- Socio-technical complexity, i.e. the interplay between a complex system and a complex organisation.

These are notably captured via the "scope", "viewpoint" and "stakeholder" attributes of SYMBIOSIS's templates. The interplay between an infrastructure and the organisation around it is a notorious cause of security issues, as illustrated by the following case study. SYMBIOSIS's templates bring such concerns to the foreground systematically.

### Illustration

The application of SYMBIOSIS to the Anthem use case gives a good illustration of the importance of comprehensive socio-technical modelling when measuring security [21]. The Anthem health insurer was robbed of nearly 80 million personal records in 2015. Hackers managed to obtain the security credentials of one or more system administrators, possibly via phishing. System administrators had access to personal data in the infrastructure, a questionable design decision, in addition to the lack of encryption of the data. The lack of understanding of the interplay between a technical system (here, a database) and members of an organisation (here, the administrator) is key in that regard.

We present now how applying SYMBIOSIS to this case would have prevented such a situation. First, consider a (partial) business objective derived from the HIPAA directive stating that only authorised personnel should have access to personal records[2], presented in Table 9.

TABLE 9
Formalised Business Objective for the Anthem case study.

| Identifier | BO2 |
|-----------|-----|
| Object | Personal records |
| Scope | Company-wide |
| Purpose | Restrict access to authorised personnel only. |
| Viewpoint | CISO |
| Context / Limitations | Permanently. |
| Relationship with other goals | Goals related to data availability requirements (not shown). |

TABLE 10
Example of Formalised Measurement Goal for the Anthem case study.

| Identifier | MG2 |
|-----------|-----|
| Object | Personal records |
| Purpose | Correctness check |
| Focus | Access rights |
| Scope | All personnel with access to personal records |
| Criteria | Access rights must be justified and timely. |
| Viewpoint | CISO, database administrators |
| Context/Limitations | Permanently. |
| Relationship with other goals | BO2 and related goals |

From objective BO2, a security measurement goal can be derived, shown in Table 10, that specifies the measurement of the validity of the access rights of any personnel with access to personal records.

From this measurement goal, a metric can be derived, shown in Table 11 (for simplicity we skip detailing the question step). The metric has immediate operational value: a monthly check will ensure that all personnel with access to personal records have up-to-date authorisations. Any violation will be reported and fixed, in collaboration with

2. HIPAA compliance [22], [19] was a requirement for health insurer Anthem.

TABLE 11
Example of a Formalised Metric for the Anthem case study.

| Identifier | ME2 |
|---|---|
| Description | Metric to ensure only authorised personnel has access to personal records. |
| Goal of measurement | MG2 |
| Base measurement | Proportion of credentials with access to personal records that have a timely and justified authorisation. |
| Measurement method | Check personal record access authorisation and expiration. |
| Measurement function | Number of credentials with justified, timely access to personal records / Number of credentials with access to personal records. |
| Measurement interpretation | 100%: no intervention required. 90-100%: notify CISO and database administrators. 0-90%: escalate to Managers responsible for data privacy and regulatory compliance for potential privacy violation and HIPAA compliance issues. |
| Reporting method and frequency | Monthly collection of base measurements, yearly analysis and reporting of selected measurements. |
| Stakeholders | CISO, personal records database administrators, data privacy manager, regulatory compliance manager. |

TABLE 12
Modelling Business Objectives in the Heartland case, highlighting the critical decomposition of scopes.

| Identifier | BO3 |
|---|---|
| Object | Cardholder data |
| Scope | Company-wide |
| Purpose | Protect (i.e. encrypt) |
| Viewpoint | CISO |
| Context / Limitations | Permanently. |
| Relationship with other goals | BO3.1, BO3.2, BO3.3. |

| Identifier | BO3.1 |
|---|---|
| Object | Cardholder data |
| Scope | Data at rest (databases) |
| Purpose | Protect (i.e. encrypt) |
| Viewpoint | CISO |
| Context / Limitations | Permanently. |
| Relationship with other goals | BO3 |

| Identifier | BO3.2 |
|---|---|
| Object | Cardholder data |
| Scope | Data in motion across the Internet |
| Purpose | Protect (i.e. encrypt) |
| Viewpoint | CISO |
| Context / Limitations | Permanently. |
| Relationship with other goals | BO3 |

| Identifier | BO3.3 |
|---|---|
| Object | Cardholder data |
| Scope | Data in motion inside the organisation's infrastructure |
| Purpose | Protect (i.e. encrypt) |
| Viewpoint | CISO |
| Context / Limitations | Permanently. |
| Relationship with other goals | BO3 |

the CISO and database administrators, and large numbers of violations will escalate the issue to data privacy and regulatory compliance departments.

### Discussion

This example extracted from the Anthem case study addresses only some of the issues that led to the original data breach, namely how to maintain consistent access control to personal records. It shows however how SYMBIOSIS integrates several orthogonal dimensions usually not considered together: regulatory compliance, privacy policies, privacy control and enforcement, and operational considerations such as "how to enforce the policy" and "what to do when a breach is detected". Different stakeholders are involved, from different domains and likely to belong to different departments, yet the templates make it explicit who is involved and in which situation: all stakeholders can trace back to the original motivation for the metric (HIPAA compliance).

#### 4.1.2 Adaptive granularity

SYMBIOSIS's breadth is a strength, yet it does not come at the cost of shallowness: the methodology is flexible and allows business objectives, measurement goals and metrics to be defined and refined at varying granularities. This flexibility is key to adapt the methodology to various organisations and domains, from large multi-national companies to small teams and individuals, while ensuring that specific details and intricacies can be captured by our models. SYMBIOSIS allows in particular to model system inter-dependencies, a common source of security breaches, as the next case study shows.

### Illustration

The Heartland case study gives a telling account of the need for well-defined security measurements at different granularities [4]. Heartland Payment Systems was breached and sensitive data was stolen despite being compliant with the Payment Card Industry Data Security Standard (PCI DSS). Applying SYMBIOSIS shows how indeed this standard was deficient as it provided incorrect, arbitrary granularities for ensuring data protection.

PCI DSS requires organisations to "Protect Cardholder Data" in the infrastructure [7]. Using SYMBIOSIS's terminology, this is a global business objective BO3 with a **scope** attribute covering "the entire infrastructure". According to PCI DSS, BO3 is to be refined into two sub-objectives (cf. Table 12):

- BO3.1: "Protecting stored cardholder data", i.e. with a scope covering data at rest in the organisation's databases.
- BO3.2: "Encrypting transmission of cardholder data across open, public networks", i.e. with a scope covering data in motion over the Internet.

By making the scope of objectives explicit, it becomes clear that this decomposition is incorrect: there should be a third objective BO3.3 with a scope covering data in motion *inside the organisation's infrastructure*. In the case of Heartland, this is indeed where attackers got a foothold and exfiltrated unencrypted data transiting inside the infrastructure.

*Discussion*

SYMBIOSIS's adaptive model that can be tailored to particular infrastructures and capture system decompositions and inter-dependencies is an important security tool. The previous example showed how incorrect scope refinement can lead to breaches. Another key aspect of the Heartland case study featured a similar incorrect decomposition: despite repeated audits and compliance certificates, the infrastructure had an undetected vulnerable public server that was compromised and used by attackers as a first penetration point. A well-defined methodology to refine security goals from global scopes to local scopes and cover an infrastructure exhaustively, such as SYMBIOSIS, is therefore key not to reproduce this kind of vulnerabilities.

### 4.2 Limitations

#### 4.2.1 Effort vs Rigour

One of the main concerns with SYMBIOSIS is the effort required for developing a clear and traceable relationship between metrics and business objectives, and also the effort associated with ensuring that the comprehensive context for metric interpretation is in place. The number of steps needed, as well as the level of detail required to fully document transitions and choices, can make the process laborious. Yet, we argue that this effort is necessary to adapt the process to the relevant granularities in the organisation, which is a key feature of the approach. As our analysis of major security incidents shows, the lack of a systematic approach often fails to unravel complex inter-dependencies leading to security failures.

#### 4.2.2 Limited support of the decision making process

SYMBIOSIS does not provide a formal decision making process that would enable stakeholders to make decisions at various transition points of the approach. As a result, the successful application of the methodology very much depends on the judgement and experience of stakeholders participating in the measurement. These stakeholders need to be mindful, during the application of SYMBIOSIS, of the methodology's tendency to grow out of control as more refinements take place and more requirements are added and intentionally try to scope and refocus it. At the same time, these stakeholders also need to ensure that transitions take place at points where subsequent steps can confidently be taken. All that being said, it is important to note that SYMBIOSIS's purpose was not to provide a means for making decisions but rather help stakeholders determine and gather information to inform their decision making process with respect to security. Notwithstanding, a more formal approach to instil confidence in transitioning between steps would be a particularly beneficial addition to the approach in future work.

## 5 CONCLUSION

In this paper, we presented SYMBIOSIS , a methodology for designing, organising and using security metrics. SYMBIOSIS allows to contextualise security metrics via the use of templates that capture relevant context elements (business goals, purpose, stakeholders, system scope). Metrics are articulated with business goals via a preliminary modelling and refining of such goals, down to a manageable granularity, while also capturing relevant context via templates. In return, this modelling allows an incremental approach, where feedback from metrics can influence business goals, and vice-versa.

We illustrated the use of SYMBIOSIS and its strengths on a set of real-world security incidents. These cases show that the lack of well-defined contextualisation methods, with a scope capturing all the necessary elements – technical, organisational, and their articulation – is key to avoid reproducing past mistakes. The flexibility of the approach allows for it to be adapted to various granularities and organisation sizes, again, an important feature considering recent security events.

In future work, we will focus on some of the shortcomings of SYMBIOSIS, including more formalised decision-support at transition points. We also plan on applying the approach to different use cases, and in particular to dynamically aggregated multi-stakeholder cyber-physical environments (such as those based on IoT) and addressing the challenge of goal and metric alignment in such settings. The multiplicity of stakeholders and the number and variety of inter-connected systems pose interesting challenges to both evaluate the effectiveness of SYMBIOSIS in such settings and develop systematic mechanisms for deriving and contextualising security metrics in such dynamically aggregated environments.

## ACKNOWLEDGEMENTS.

This work is supported by UK Engineering and Physical Sciences Research Council (EPSRC) Grant Mumba: Multi-faceted Metrics for ICS Business Risk Analysis (EP/M002780/1), part of the UK Research Institute on Trustworthy Industrial Control Systems (RITICS). Awais Rashid's research in cyber security risk and metrics is also supported by an Alan Turing Institute Fellowship.